\documentclass[cameraready]{Interspeech}

\title{Disentangling Depression from Cognitive Decline in Elderly Speech \\Using Concurrent Clinical Assessments}

\author[affiliation={1}]{Woori}{Jeon}
\author[affiliation={2}]{Seunghee}{Ha}
\author[affiliation={2}]{Sang-Kyu}{Lee}
\author[affiliation={2}]{Ji Hye}{Yoon}
\author[affiliation={3}]{Tae-Jin}{Yoon}
\author[affiliation={2}]{Seung Jin}{Lee}
\author[affiliation={2}]{Woojae}{Han}
\author[affiliation={1}, correspondingauthor]{Jungmin}{So}

\address{
    $^1$ Sogang University, South Korea \\
    $^2$ Hallym University, South Korea \\
    $^3$ Sungshin Women's University, South Korea
}

\email{woori50300@gmail.com, shha@hallym.ac.kr, skmind@hallym.ac.kr, j.yoon@hallym.ac.kr, tyoon@sungshin.ac.kr, sjl@hallym.ac.kr, woojaehan@hallym.ac.kr, jso1@sogang.ac.kr}

\keywords{depression detection, elderly speech, mild cognitive impairment, concurrent clinical assessments}

\usepackage{multirow}
\usepackage{booktabs}
\usepackage{graphicx}

\makeatletter
\def\makeLineNumber{%
  \if@firstcolumn
    \makeLineNumberLeft
  \else
    \makeLineNumberRight
  \fi
}
\makeatother

\begin{document}

\maketitle

\begin{abstract}
Detecting depression from speech in elderly patients with mild cognitive impairment (MCI) is complicated by overlapping acoustic effects of cognitive decline. Without disentangling these, classifiers risk learning cognitive rather than depression-specific patterns. We present a Korean elderly speech corpus of 89 MCI speakers collected over three years with concurrent depression (SGDS) and cognitive (MMSE) assessments. Both linear mixed-effects models and partial correlations identify the same pattern: of seven feature groups, only formant features are associated with depression after controlling for cognitive function, while widely used F0 features carry no signal. The F1+F2 subset achieves UAR 0.760 with just 12 features, outperforming both the full eGeMAPS baseline and self-supervised representations. These findings show that concurrent clinical assessments can identify interpretable, depression-specific acoustic markers in populations where cognitive and affective symptoms co-occur.
\end{abstract}

\section{Introduction}

Depression is highly prevalent among elderly individuals with mild cognitive impairment (MCI), affecting approximately 32\% \cite{ismail2017prevalence} compared to 7\% in the general elderly population \cite{alexopoulos2005depression}. Accurate detection is clinically critical because untreated depression accelerates cognitive decline. Speech-based screening offers a promising non-invasive approach \cite{cummins2015review, low2020automated}, achieving strong performance on healthy adults using corpora such as DAIC-WOZ \cite{gratch2014distress} with handcrafted features \cite{deng2025interpretable} and self-supervised representations \cite{wu2023ssl}. Recent studies have further explored longitudinal monitoring \cite{lewis2025longitudinal, hidalgo2025biomarkers}, foundation models \cite{gomezzaragoza2025foundation, white2025reference}, domain adaptation \cite{dumpala2025ttt}, multimodal fusion \cite{you2025m3l}, and ordinal regression \cite{zuo2025ordinal}.

However, a fundamental challenge arises in MCI populations: both depression and cognitive decline produce overlapping acoustic changes such as reduced pitch range, decreased vocal energy, and altered voice quality \cite{cannizzaro2004voice, konig2015automatic}, so classifiers may learn cognitive patterns rather than depression-specific ones. A recent systematic review confirms that insufficient control of cognitive confounders remains a key limitation in speech-based depression detection for older adults \cite{martinez2025speech}.

In general adult populations, F0 features such as pitch range and variability are among the most widely studied acoustic markers of depression \cite{cummins2015review}, while formant and vowel space features have also been identified as robust indicators: reduced vowel space characterizes depressed speech across multiple corpora \cite{scherer2015vowel}, gender-dependent vowel-level formant features outperform full feature sets \cite{vlasenko2017gender}, and formant features show significant group differences between depressed and healthy speakers \cite{menne2024voice}. More broadly, depression reduces acoustic space variability, reflecting decreased articulatory effort \cite{cummins2015acoustic}. However, cognitive decline also constrains articulatory dynamics, making it unclear which of these changes observed in MCI populations reflect depression and which reflect cognitive decline.

Resolving this ambiguity requires concurrent depression and cognitive scores at each assessment, enabling identification of which features are specifically associated with depression after accounting for cognitive function. Young et al.\ \cite{young2025comorbid} proposed feature weighting to mitigate depression--dementia confounding in elderly speech, but without concurrent cognitive scores to identify which specific features are depression-specific vs.\ cognition-related. No prior study has used dual clinical assessments to systematically separate depression-specific from cognition-related acoustic features in MCI elderly speech.

In this paper, we make the following contributions:
\begin{enumerate}
    \item We introduce the first Korean elderly MCI speech corpus: 89 individuals recorded annually over three years (209 observations), with concurrent depression (SGDS) and cognitive (MMSE) scores at every recording. Within-subject depression status changes across years enable longitudinal analysis of depression-specific acoustic effects in cognitively impaired populations.
    \item Two independent methods---longitudinal LMMs and cross-sectional partial correlations---both identify formant as the only depression-specific feature group among seven eGeMAPS groups after controlling for cognitive function, while widely used F0 features carry no signal. This convergence across methods provides evidence robust to methodological choice.
    \item The statistically identified formant group (18 features) achieves UAR 0.747, and the F1+F2 subset further improves to 0.760 with just 12 interpretable features, outperforming both the full 88-feature eGeMAPS baseline and self-supervised representations.
\end{enumerate}

\section{Data}

\subsection{Corpus}

We collected a longitudinal Korean speech corpus from elderly individuals clinically diagnosed with MCI by a psychiatrist at a university hospital. The study was approved by the institutional review board. Participants were recorded annually over three years. Of 89 participants who were recorded at least once, 72 participated in two or more years, yielding 209 total observations.

Table~\ref{tab:data} summarizes the corpus statistics. All participants were community-dwelling elderly aged 60 or older, with a mean age of 78.1 years (SD 6.2), and diagnosed with MCI. 67\% were female, with a mean education of 8.1 years. Depression was assessed annually using the Short-form Geriatric Depression Scale (SGDS) \cite{yesavage1983gds, bae1999sgds}, with scores $\geq 8$ indicating depression. Mean SGDS scores ranged from 4.0 to 4.6 across years, and 41 of 209 observations fell above the clinical threshold.

Importantly, 18 of 89 subjects changed depression status across years, with 11 transitions from non-depressed to depressed and 10 from depressed to non-depressed. This within-subject variation enables the LMM analysis in Section~3, where fixed effects of SGDS and MMSE are estimated from within-subject change over time. Cognitive function was measured by the Mini-Mental State Examination (MMSE) \cite{folstein1975mmse}, which remained relatively stable across years.

Each recording session comprises four speech tasks---passage reading, sustained vowel phonation, sentence repetition, and a structured depressive symptom interview---along with additional clinical measures including anxiety, quality of life, and perceived stress. This study analyzes the passage reading task using depression and cognitive scores; the remaining tasks and clinical measures are available for future research.

\begin{table}[t]
  \caption{Corpus statistics. Dep.\ = depressed (SGDS $\geq$ 8).}
  \label{tab:data}
  \centering
  \footnotesize
  \begin{tabular}{@{}l c cc c c c@{}}
    \toprule
    \textbf{Year} & \textbf{N} & \textbf{Dep.} & \textbf{Non-d.} & \textbf{Age} & \textbf{MMSE} & \textbf{SGDS} \\
    \midrule
    Y1 & 61 & 14 & 47 & 77.7$\pm$5.8 & 23.8$\pm$3.3 & 4.5$\pm$3.9 \\
    Y2 & 72 & 13 & 59 & 78.5$\pm$6.1 & 23.2$\pm$3.5 & 4.0$\pm$4.1 \\
    Y3 & 76 & 14 & 62 & 78.1$\pm$6.5 & 23.4$\pm$3.2 & 4.6$\pm$3.3 \\
    \midrule
    Total & 209 & 41 & 168 & 78.1$\pm$6.2 & 23.5$\pm$3.3 & 4.4$\pm$3.8 \\
    \bottomrule
  \end{tabular}
\end{table}

\subsection{Speech task}

Participants read aloud the ``Autumn'' passage \cite{kim2012neurologic}, a standardized Korean reading passage used in clinical speech assessment, in a quiet room at a university hospital. The passage contains approximately 210 syllables and takes 30--60 seconds to read aloud. Recordings were captured at 48\,kHz, 16-bit with an 8-channel array and downmixed to mono for analysis. This reading task was chosen for two reasons. First, it controls linguistic content across participants, ensuring that acoustic differences reflect vocal rather than linguistic variation, which is critical when using phonetically sensitive features such as eGeMAPS. Second, reading reduces cognitive load compared to spontaneous speech, avoiding the risk that high cognitive demands during free speech confound acoustic measurements with task difficulty effects in MCI populations \cite{konig2015automatic}.
\subsection{Feature extraction}

Examiner instructions at the beginning of each recording were identified using WhisperX \cite{bain2023whisperx} and removed prior to feature extraction. We extract the extended Geneva Minimalistic Acoustic Parameter Set (eGeMAPS) \cite{eyben2016opensmile}, yielding 88 interpretable acoustic features spanning seven groups: fundamental frequency (F0, 10 features), energy/loudness (12), spectral balance (17), MFCC (16), voice quality (10), formant (18), and temporal (5). Additionally, we compute character error rate (CER) and word error rate (WER) by running Whisper large-v3 \cite{radford2023whisper} on each recording and comparing against the original passage text, serving as objective measures of speech intelligibility.

\section{Methods}

In populations with concurrent depression and cognitive decline, standard classification risks conflating the two conditions in the learned representation. Our approach therefore proceeds in two stages. First, we use statistical analysis to identify which acoustic feature groups are specifically associated with depression after controlling for cognitive function (Sections~3.1--3.2). Second, we evaluate whether these statistically identified groups also yield better classification performance (Section~3.3).

\subsection{Group-level feature representation}

Testing 88 individual features introduces a substantial multiple comparison burden and yields results that are difficult to interpret clinically. We therefore analyze at the group level, where each group corresponds to an acoustically coherent family of features. Based on the parameter categories in \cite{eyben2016opensmile}, we organize the 88 features into seven functionally motivated groups (Table~\ref{tab:groups}). These groups are defined before any classification experiment, ensuring that group definitions are independent of classification outcomes. This reduces the number of statistical tests from 88 to 7 while producing interpretable conclusions such as ``formant features are associated with depression'' rather than lists of individual feature names.

\begin{table}[t]
\centering
\caption{Seven feature groups derived from the 88 eGeMAPS features. Each LLD is expanded into mean and standard deviation over voiced (V), unvoiced (UV), or all frames, yielding the total count per group. Temporal features are already computed per utterance, so no mean/standard deviation is applied.}
\label{tab:groups}
\setlength{\tabcolsep}{6pt}
\begin{tabular}{llr}
\toprule
\textbf{Group} & \textbf{Low-level descriptors} & \textbf{\#} \\
\midrule
F0          & F0 (semitone), rising/falling slope       & 10 \\
Energy      & Loudness, peaks/sec, equiv.\ sound level  & 12 \\
Spectral    & Alpha ratio, Hammarberg idx, slope,       & 17 \\
            & spectral flux (V/UV)                      &    \\
MFCC        & MFCC 1--4 (all frames + V)                & 16 \\
Voice qual. & Jitter, shimmer, HNR, H1--H2, H1--A3     & 10 \\
Formant     & F1--F3 frequency, bandwidth, rel.\ energy & 18 \\
Temporal    & Voiced/unvoiced segment rate \& duration   &  5 \\
\midrule
\multicolumn{2}{l}{Total}                               & 88 \\
\bottomrule
\end{tabular}
\end{table}

For each group, we standardize its constituent features to zero mean and unit variance, then extract the first principal component (PC1) via PCA. PC1 captures the dominant pattern of covariation within each group and serves as the dependent variable in the statistical analyses described in Section~3.2.

\subsection{Disentangling depression from cognitive effects}

To ensure that the identified associations are not artifacts of a single statistical framework, we apply two independent methods to each group's PC1: one capturing within-subject change over time, and one capturing between-subject differences at each time point.

\textbf{Linear mixed-effects model (LMM).} We fit the following LMM \cite{bates2015lme4} for each group's PC1:
\begin{align}
  y \sim \text{SGDS} + \text{MMSE} + \text{year}_c + \text{female} + \text{age} + (1|\text{subj})
  \label{eq:lmm}
\end{align}
where $y$ is a group-level PC1, $\text{year}_c$ is year centered at the middle collection year, $\text{female}$ is a binary gender indicator, $\text{age}$ is age at recording, and $(1|\text{subj})$ is a random intercept per subject. The random intercept absorbs all time-invariant between-subject variability (e.g., vocal tract anatomy, habitual speaking style), so the fixed effects of SGDS and MMSE capture only \textit{within-subject} associations between clinical scores and acoustic change. We use a random intercept rather than random slope model because many subjects have only two observations, insufficient for slope estimation. All features are $z$-scored across all observations before model fitting.

By including both SGDS and MMSE as fixed effects, the $\beta_\text{SGDS}$ coefficient estimates each group's association with depression severity after accounting for cognitive function. This isolates depression-related acoustic variation from changes that may reflect cognitive decline, a critical distinction for MCI populations where the two conditions frequently co-occur.

\textbf{Partial correlation.} As the between-subject counterpart, we compute partial Spearman correlations between each group's PC1 and SGDS, controlling for MMSE, age, and gender:
\begin{align}
  e_{\text{PC1}} &= \text{PC1} - \mathbf{Z}\hat{\boldsymbol{\gamma}}_1, \quad
  e_{\text{SGDS}} = \text{SGDS} - \mathbf{Z}\hat{\boldsymbol{\gamma}}_2 \notag \\
  r_s &= \text{Spearman}(e_{\text{PC1}},\; e_{\text{SGDS}})
  \label{eq:pcorr}
\end{align}
where $\mathbf{Z} = [\text{MMSE}, \text{age}, \text{gender}]$ is the covariate matrix and $\hat{\boldsymbol{\gamma}}$ are ordinary least-squares coefficients.

Convergence between these two methodologically independent approaches (one longitudinal, one cross-sectional) provides stronger evidence that the identified association is robust and not an artifact of either statistical framework.

\subsection{Classification}

For binary depression classification (SGDS $\geq 8$), we use a support vector machine (SVM) with RBF kernel and $C = 0.1$, a standard classifier for small-sample paralinguistic tasks. Each feature group is evaluated using all its constituent features (not PC1), so that the statistical analysis identifies \textit{which} group to use while the classifier exploits the full feature space within that group. Class weights are set inversely proportional to class frequencies to address the label imbalance (41 depressed vs.\ 168 non-depressed). Features are standardized within each cross-validation fold: the scaler is fit on training data only and applied to both training and test sets to prevent information leakage.

Evaluation uses leave-one-subject-out (LOSO) cross-validation on all 209 observations (89 subjects), where all observations from the same subject are held out together. We report unweighted average recall (UAR), sensitivity, specificity, accuracy, and F1 score.

We compare two handcrafted baselines: (1)~full eGeMAPS (88 features), representing the standard feature set, and (2)~CER + WER from Whisper large-v3 (2 features), representing speech intelligibility measures. As self-supervised baselines, we evaluate four models---WavLM-Base \cite{chen2022wavlm}, wav2vec2-Base \cite{baevski2020wav2vec2}, HuBERT-Base \cite{hsu2021hubert}, and WavLM-Large---using mean-pooled embeddings with both SVM and a linear classification head. Linear head results are averaged over 5 random seeds for stability.

\section{Results}

\subsection{Disentangling analysis}\label{sec:disentangling}

Table~\ref{tab:lmm} presents the group-level results from both the LMM and partial correlation. Of the seven feature groups, formant is the only one that reaches significance in both methods: $\beta_\text{SGDS} = -0.095$, $p = 0.011$ in the LMM, and $r = -0.196$, $p = 0.004$ in the partial correlation. The negative coefficients indicate that greater depression severity is associated with a systematic shift in the formant pattern. Energy and spectral also reach LMM significance with $p = 0.023$ and $p = 0.028$, but neither is confirmed by the partial correlation, suggesting these groups may capture within-subject changes that do not generalize across individuals. Because both methods control for MMSE, these associations reflect depression severity independent of cognitive function. The remaining four groups, F0, MFCC, voice quality, and temporal, show no significant association with depression in either method.

The formant partial correlation survives Bonferroni correction for the seven group tests ($\alpha_\text{adj} = 0.05/7 = 0.007$), and the convergence of two independent statistical frameworks on the same group strengthens the evidence that this association is robust.

F0 is one of the most widely used features in speech-based depression detection, yet it shows no association with depression in this MCI population, with $p = 0.894$ in the LMM and $p = 0.478$ in the partial correlation.

\begin{table}[t]
  \caption{Group-level analysis: LMM SGDS coefficient and partial correlation with SGDS controlling for MMSE, age, and gender, applied to PC1 of each group. $^{*}p<.05$.}
  \label{tab:lmm}
  \centering
  \footnotesize
  \begin{tabular}{@{}l r r rr rr@{}}
    \toprule
    & & & \multicolumn{2}{c}{\textbf{LMM}} & \multicolumn{2}{c}{\textbf{Pcorr}} \\
    \cmidrule(lr){4-5} \cmidrule(lr){6-7}
    \textbf{Group} & $d$ & \textbf{\%PC1} & $\beta$ & $p$ & $r$ & $p$ \\
    \midrule
    F0 & 10 & 40 & $+.005$ & $.894$ & $+.049$ & $.478$ \\
    MFCC & 16 & 24 & $+.025$ & $.432$ & $+.092$ & $.188$ \\
    Temporal & 5 & 54 & $-.034$ & $.281$ & $-.127$ & $.066$ \\
    Voice qual. & 10 & 38 & $+.009$ & $.822$ & $-.039$ & $.573$ \\
    Energy & 12 & 74 & $-.097$ & $.023^*$ & $-.061$ & $.384$ \\
    Spectral & 17 & 27 & $+.079$ & $.028^*$ & $+.074$ & $.289$ \\
    \textbf{Formant} & \textbf{18} & \textbf{30} & $\mathbf{-.095}$ & $\mathbf{.011}^*$ & $\mathbf{-.196}$ & $\mathbf{.004}^*$ \\
    \bottomrule
  \end{tabular}
\end{table}

\subsection{Classification}

Table~\ref{tab:classification} presents the classification results. The formant group achieves the highest UAR of 0.747 with 18 features, outperforming both the 88-feature eGeMAPS baseline at 0.655 and the best SSL model, HuBERT with a linear head, at 0.729.

F0 yields a UAR of 0.512, essentially at chance, confirming that these features carry no useful signal for depression in this population. The remaining groups fall between these extremes (UAR 0.585--0.641). The three groups that were LMM-significant in Table~\ref{tab:lmm} are also the three best classifiers, while all four non-significant groups score below 0.600. This correspondence suggests that the statistical analysis in Section~4.1 successfully identifies the groups carrying genuine depression signal.

By contrast, selecting the top 18 individual features by partial correlation or LMM, matching the dimensionality of the formant group, yields UAR of 0.655 and 0.637 respectively, no better than the full baseline. Individual features drawn from heterogeneous groups lack the coherent signal that a functionally related group provides, confirming the advantage of group-level selection.

CER + WER achieves 0.645 with just two features, matching the 88-feature baseline and illustrating the curse of dimensionality with a small sample. Among SSL models, linear classification heads consistently outperform SVM across all four architectures. HuBERT improves from 0.691 with SVM to 0.729 with a linear head, suggesting that the non-linear SVM kernel may overfit in high-dimensional space with limited training data. The formant group outperforms all SSL models with only 18 statistically grounded features.

\begin{table}[t]
  \caption{Depression classification (LOSO CV, $N$=209). $^\dagger$Linear head (5-seed average); all others SVM RBF $C$=0.1. Best UAR in bold.}
  \label{tab:classification}
  \centering
  \footnotesize
  \begin{tabular*}{\columnwidth}{@{\extracolsep{\fill}}l r ccccc@{}}
    \toprule
    \textbf{Method} & $d$ & \textbf{UAR} & \textbf{Sens} & \textbf{Spec} & \textbf{Acc} & \textbf{F1} \\
    \midrule
    \multicolumn{7}{@{}l}{\textit{Baselines}} \\
    eGeMAPS full & 88 & .655 & .512 & .798 & .742 & .438 \\
    CER + WER & 2 & .645 & .683 & .607 & .622 & .415 \\
    \midrule
    \multicolumn{7}{@{}l}{\textit{SSL + SVM}} \\
    WavLM-Base & 768 & .662 & .634 & .690 & .679 & .437 \\
    wav2vec2-Base & 768 & .674 & .585 & .762 & .727 & .457 \\
    HuBERT-Base & 768 & .691 & .537 & .845 & .785 & .494 \\
    WavLM-Large & 1024 & .692 & .659 & .726 & .713 & .474 \\
    \midrule
    \multicolumn{7}{@{}l}{\textit{SSL + linear$^\dagger$}} \\
    WavLM-Base & 768 & .696 & .766 & .626 & .654 & .464 \\
    wav2vec2-Base & 768 & .681 & .717 & .644 & .658 & .452 \\
    HuBERT-Base & 768 & .729 & .795 & .663 & .689 & .501 \\
    WavLM-Large & 1024 & .705 & .776 & .635 & .662 & .474 \\
    \midrule
    \multicolumn{7}{@{}l}{\textit{Individual selection (top-18)}} \\
    Pcorr top-18 & 18 & .655 & .561 & .750 & .713 & .434 \\
    LMM top-18 & 18 & .637 & .512 & .762 & .713 & .412 \\
    \midrule
    \multicolumn{7}{@{}l}{\textit{Feature groups}} \\
    F0 & 10 & .512 & .537 & .488 & .498 & .295 \\
    MFCC & 16 & .585 & .390 & .780 & .703 & .340 \\
    Temporal & 5 & .588 & .634 & .542 & .560 & .361 \\
    Voice quality & 10 & .599 & .317 & .881 & .770 & .351 \\
    Energy & 12 & .638 & .561 & .714 & .684 & .411 \\
    Spectral & 17 & .641 & .561 & .720 & .689 & .414 \\
    \textbf{Formant} & 18 & \textbf{.747} & .756 & .738 & .742 & .534 \\
    \bottomrule
  \end{tabular*}
\end{table}

\subsection{Formant subgroup analysis}

We next examine which formants drive the formant group's performance. Each eGeMAPS formant contributes six features: mean and coefficient of variation for frequency, bandwidth, and amplitude. Table~\ref{tab:formant_sub} reports all single- and pairwise combinations.

F2 alone matches the full formant group at UAR 0.747 with only 6 features, while F1 at 0.618 and F3 at 0.641 perform considerably worse. Adding F1 to F2 yields the highest UAR of 0.760, whereas adding F3 to any F2-containing subset degrades performance. This indicates that F2 carries the primary depression signal and F1 provides complementary information, while F3 introduces noise. F1 and F2 are the primary acoustic correlates of tongue position and vowel quality, suggesting that depression in this population is associated with articulatory changes. F3, which primarily reflects lip rounding, does not contribute to this signal. Permutation tests confirm that the F1+F2 advantage is significant over both the full eGeMAPS baseline ($p = 0.037$) and individual feature selection ($p = 0.007$ for Pcorr, $p = 0.008$ for LMM top-18).

\begin{table}[t]
  \caption{Formant subgroup classification. Same setup as Table~\ref{tab:classification}. Best UAR in bold.}
  \label{tab:formant_sub}
  \centering
  \footnotesize
  \begin{tabular*}{\columnwidth}{@{\extracolsep{\fill}}l r ccccc@{}}
    \toprule
    \textbf{Subgroup} & $d$ & \textbf{UAR} & \textbf{Sens} & \textbf{Spec} & \textbf{Acc} & \textbf{F1} \\
    \midrule
    F1 & 6 & .618 & .634 & .601 & .608 & .388 \\
    F2 & 6 & .747 & .756 & .738 & .742 & .534 \\
    F3 & 6 & .641 & .585 & .696 & .675 & .414 \\
    \midrule
    \textbf{F1+F2} & 12 & \textbf{.760} & .805 & .714 & .732 & .541 \\
    F1+F3 & 12 & .662 & .634 & .690 & .679 & .437 \\
    F2+F3 & 12 & .735 & .707 & .762 & .751 & .527 \\
    \midrule
    F1+F2+F3 & 18 & .747 & .756 & .738 & .742 & .534 \\
    \bottomrule
  \end{tabular*}
\end{table}

\section{Discussion}

\subsection{Error analysis and interpretation}

Three sources of variation beyond depression affect acoustic features in this population: gender, aging, and cognition.

Error analysis of the F1+F2 model reveals a gender asymmetry: 43 of 48 false positives are female, yielding male specificity of 0.911 versus female specificity of 0.616. Gender is a major source of formant variance, and in a gender-pooled model this variation can be confounded with depression-related formant changes, disproportionately affecting females who constitute 67\% of our sample. Vlasenko et al.\ \cite{vlasenko2017gender} similarly found that gender-dependent formant features outperform gender-pooled ones, and gender-specific normalization could reduce this bias.

The absence of F0 signal may reflect age-related F0 variation overwhelming clinical signal in elderly populations \cite{konig2015automatic}, implying that depression models trained on younger adults may fail for elderly MCI populations. A sensitivity analysis confirms that controlling for age and gender alone is insufficient: temporal features lose significance only when MMSE is added ($p$: 0.004 $\to$ 0.066), while formant features remain significant across all control levels ($p = 0.004$). Concurrent cognitive assessment is therefore essential for feature selection in MCI populations.

\subsection{Limitations}

Several limitations should be noted. The sample size of 89 subjects and 209 observations is modest. The formant partial correlation survives Bonferroni correction ($p = 0.004$), but the LMM $p$-value does not, and replication on larger cohorts is needed. The gender imbalance of 67\% female and the absence of gender-specific modeling limit specificity for female participants. We evaluate only a reading task; spontaneous speech, which better reflects naturalistic communication patterns, could yield complementary markers. Finally, cross-linguistic validation is needed, as our corpus is Korean only.

\section{Conclusion}

In this paper, we showed that concurrent clinical assessments enable disentangling depression from cognitive decline in elderly MCI speech, revealing formant as the only depression-specific feature group. We achieve UAR 0.760 with just 12 F1+F2 features, outperforming both the full eGeMAPS baseline and self-supervised representations. Future work includes gender-specific modeling, spontaneous speech analysis, and longitudinal tracking of acoustic change leveraging the within-subject variation in our corpus.

\ifcameraready
\section{Acknowledgments}
This work was supported by the Ministry of Education of the Republic of Korea and the National Research Foundation of Korea (NRF-2025S1A5C3A02005633), by the Artificial Intelligence Innovation Graduate School (Sogang University) grant funded by the Korea government (MSIT) (RS-2026-25547954), and by the National Information Society Agency (NIA) funded by the Ministry of Science, ICT through the Big Data Platform and Center Construction Project (No. 2022-DATA-W18).
\fi

\section{Generative AI Use Disclosure}
Generative AI tools (Claude) were used for code development assistance and manuscript editing. All analyses, experimental design, and scientific conclusions were made by the authors.

\bibliographystyle{IEEEtran}
\bibliography{mybib}

@article{cummins2015review,
  author    = {Cummins, Nicholas and Scherer, Stefan and Krajewski, Jarek and Schnieder, Sebastian and Epps, Julien and Quatieri, Thomas F.},
  title     = {A review of depression and suicide risk assessment using speech analysis},
  journal   = {Speech Communication},
  volume    = {71},
  pages     = {10--49},
  year      = {2015},
  publisher = {Elsevier}
}

@inproceedings{gratch2014distress,
  author    = {Gratch, Jonathan and Artstein, Ron and Lucas, Gale M. and Stratou, Giota and Scherer, Stefan and Nazarian, Angela and Wood, Rachel and Boberg, Jill and DeVault, David and Marsella, Stacy and Traum, David and Rizzo, Albert and Morency, Louis-Philippe},
  title     = {The Distress Analysis Interview Corpus of human and computer interviews},
  booktitle = {Proc. LREC},
  pages     = {3123--3128},
  year      = {2014}
}

@inproceedings{young2025comorbid,
  author    = {Young, Sophie and Tao, Fuxiang and Mirheidari, Bahman and Pahar, Madhurananda and Reuber, Markus and Christensen, Heidi},
  title     = {Can Speech Accurately Detect Depression in Patients With Comorbid Dementia? {An} Approach for Mitigating Confounding Effects of Depression and Dementia},
  booktitle = {Proc. Interspeech},
  year      = {2025},
  address   = {Rotterdam, Netherlands}
}

@inproceedings{lewis2025longitudinal,
  author    = {Lewis, Robert and Fedor, Szymon and Hidalgo Julia, Nelson and Kim, Jiyeon and Jones, Noah and others},
  title     = {Towards the Objective Characterisation of Major Depressive Disorder Using Speech Data from a 12-week Observational Study with Daily Measurements},
  booktitle = {Proc. Interspeech},
  year      = {2025},
  address   = {Rotterdam, Netherlands}
}

@inproceedings{hidalgo2025biomarkers,
  author    = {Hidalgo Julia, Nelson and Lewis, Robert and Ferguson, C. and Goldberg, S. and Lau, W. and others},
  title     = {Identifying Vocal and Facial Biomarkers of Depression in Large-Scale Remote Recordings: {A} Multimodal Study Using Mixed-Effects Modeling},
  booktitle = {Proc. Interspeech},
  year      = {2025},
  address   = {Rotterdam, Netherlands}
}

@inproceedings{white2025reference,
  author    = {White, Lauren and Carr, Ewan and Dineley, Judith and Botelho, Catarina and Conde, Pauline and others},
  title     = {Speech Reference Intervals: {An} Assessment of Feasibility in Depression Symptom Severity Prediction},
  booktitle = {Proc. Interspeech},
  year      = {2025},
  address   = {Rotterdam, Netherlands}
}

@inproceedings{deng2025interpretable,
  author    = {Deng, Qingkun and Luz, Saturnino and de la Fuente Garcia, Sofia},
  title     = {An Interpretable Speech Foundation Model for Depression Detection by Revealing Prediction-Relevant Acoustic Features from Long Speech},
  booktitle = {Proc. Interspeech},
  year      = {2025},
  address   = {Rotterdam, Netherlands}
}

@inproceedings{wu2023ssl,
  author    = {Wu, Wen and Zhang, Chao and Woodland, Philip C.},
  title     = {Self-Supervised Representations in Speech-Based Depression Detection},
  booktitle = {Proc. ICASSP},
  pages     = {1--5},
  year      = {2023},
  organization = {IEEE}
}

@article{eyben2016opensmile,
  author    = {Eyben, Florian and Scherer, Klaus R. and Schuller, Bj{\"o}rn W. and Sundberg, Johan and Andr{\'e}, Elisabeth and Busso, Carlos and Devillers, Laurence and Epps, Julien and Laukka, Patrik and Narayanan, Shrikanth S. and Truong, Khiet P.},
  title     = {The {Geneva} Minimalistic Acoustic Parameter Set ({GeMAPS}) for Voice Research and Affective Computing},
  journal   = {IEEE Transactions on Affective Computing},
  volume    = {7},
  number    = {2},
  pages     = {190--202},
  year      = {2016}
}

@inproceedings{bain2023whisperx,
  author    = {Bain, Max and Huh, Jaesung and Han, Tengda and Zisserman, Andrew},
  title     = {{WhisperX}: Time-Accurate Speech Transcription of Long-Form Audio},
  booktitle = {Proc. Interspeech},
  year      = {2023}
}

@inproceedings{radford2023whisper,
  author    = {Radford, Alec and Kim, Jong Wook and Xu, Tao and Brockman, Greg and McLeavey, Christine and Sutskever, Ilya},
  title     = {Robust Speech Recognition via Large-Scale Weak Supervision},
  booktitle = {Proc. ICML},
  pages     = {28492--28518},
  year      = {2023}
}

@article{yesavage1983gds,
  author    = {Yesavage, Jerome A. and Brink, T. L. and Rose, Terence L. and Lum, Owen and Huang, Virginia and Adey, Michael and Leirer, Von Otto},
  title     = {Development and validation of a geriatric depression screening scale: a preliminary report},
  journal   = {Journal of Psychiatric Research},
  volume    = {17},
  number    = {1},
  pages     = {37--49},
  year      = {1983}
}

@article{folstein1975mmse,
  author    = {Folstein, Marshal F. and Folstein, Susan E. and McHugh, Paul R.},
  title     = {{``Mini-mental state''}. {A} practical method for grading the cognitive state of patients for the clinician},
  journal   = {Journal of Psychiatric Research},
  volume    = {12},
  number    = {3},
  pages     = {189--198},
  year      = {1975}
}

@article{bae1999sgds,
  author    = {Bae, Jong Nim and Cho, Maeng Je},
  title     = {Development of the {Korean} version of the {Geriatric Depression Scale} and its short form among elderly psychiatric patients},
  journal   = {Journal of Psychosomatic Research},
  volume    = {57},
  number    = {3},
  pages     = {297--305},
  year      = {2004}
}

@article{low2020automated,
  author    = {Low, Daniel M. and Bentley, Kate H. and Ghosh, Satrajit S.},
  title     = {Automated assessment of psychiatric disorders using speech: A systematic review},
  journal   = {Laryngoscope Investigative Otolaryngology},
  volume    = {5},
  number    = {1},
  pages     = {96--116},
  year      = {2020}
}

@article{cannizzaro2004voice,
  author    = {Cannizzaro, Michael and Harel, Brian and Reilly, Nicole and Chappell, Phillip and Snyder, Peter J.},
  title     = {Voice acoustical measurement of the severity of major depression},
  journal   = {Brain and Cognition},
  volume    = {56},
  number    = {1},
  pages     = {30--35},
  year      = {2004}
}

@article{konig2015automatic,
  author    = {K{\"o}nig, Alexandra and Satt, Aharon and Sorin, Alexander and Hoory, Ron and Toledo-Ronen, Orith and Derreumaux, Alexandre and Manera, Valeria and Verhey, Frans and Aalten, Pauline and Robert, Philippe H. and David, Renaud},
  title     = {Automatic speech analysis for the assessment of patients with predementia and {Alzheimer's} disease},
  journal   = {Alzheimer's \& Dementia: Diagnosis, Assessment \& Disease Monitoring},
  volume    = {1},
  number    = {1},
  pages     = {112--124},
  year      = {2015}
}

@article{bates2015lme4,
  author    = {Bates, Douglas and M{\"a}chler, Martin and Bolker, Ben and Walker, Steve},
  title     = {Fitting Linear Mixed-Effects Models Using lme4},
  journal   = {Journal of Statistical Software},
  volume    = {67},
  number    = {1},
  pages     = {1--48},
  year      = {2015}
}

@article{alexopoulos2005depression,
  author    = {Alexopoulos, George S.},
  title     = {Depression in the elderly},
  journal   = {The Lancet},
  volume    = {365},
  number    = {9475},
  pages     = {1961--1970},
  year      = {2005}
}

@article{ismail2017prevalence,
  author    = {Ismail, Zahinoor and Elbayoumi, Heba and Fischer, Corinne E. and Hogan, David B. and Bherer, Louis and others},
  title     = {Prevalence of depression in patients with mild cognitive impairment: a systematic review and meta-analysis},
  journal   = {JAMA Psychiatry},
  volume    = {74},
  number    = {1},
  pages     = {58--67},
  year      = {2017}
}

@inproceedings{scherer2015vowel,
  author    = {Scherer, Stefan and Morency, Louis-Philippe and Gratch, Jonathan and Pestian, John},
  title     = {Reduced Vowel Space is a Robust Indicator of Psychological Distress: A Cross-Corpus Analysis},
  booktitle = {Proc. IEEE ICASSP},
  pages     = {4789--4793},
  year      = {2015}
}

@inproceedings{vlasenko2017gender,
  author    = {Vlasenko, Bogdan and Sagha, Hesam and Cummins, Nicholas and Schuller, Bj{\"o}rn},
  title     = {Implementing Gender-Dependent Vowel-Level Analysis for Boosting Speech-Based Depression Recognition},
  booktitle = {Proc. Interspeech},
  pages     = {3266--3270},
  year      = {2017}
}

@article{cummins2015acoustic,
  author    = {Cummins, Nicholas and Sethu, Vidhyasaharan and Epps, Julien and Schnieder, Sebastian and Krajewski, Jarek},
  title     = {Analysis of acoustic space variability in speech affected by depression},
  journal   = {Speech Communication},
  volume    = {75},
  pages     = {27--49},
  year      = {2015}
}

@article{martinez2025speech,
  author    = {Martinez-Nicolas, Israel and Criado, Diego and Gordillo, Fernando and Martinez-Sanchez, Francisco and Meilan, Juan Jose G.},
  title     = {Speech analysis for detecting depression in older adults: a systematic review},
  journal   = {Frontiers in Psychology},
  volume    = {16},
  pages     = {1715538},
  year      = {2025}
}

@inproceedings{baevski2020wav2vec2,
  author    = {Baevski, Alexei and Zhou, Yuhao and Mohamed, Abdelrahman and Auli, Michael},
  title     = {wav2vec 2.0: A Framework for Self-Supervised Learning of Speech Representations},
  booktitle = {Proc. NeurIPS},
  volume    = {33},
  pages     = {12449--12460},
  year      = {2020}
}

@article{hsu2021hubert,
  author    = {Hsu, Wei-Ning and Bolte, Benjamin and Tsai, Yao-Hung Hubert and Lakhotia, Kushal and Salakhutdinov, Ruslan and Mohamed, Abdelrahman},
  title     = {{HuBERT}: Self-Supervised Speech Representation Learning by Masked Prediction of Hidden Units},
  journal   = {IEEE/ACM Transactions on Audio, Speech, and Language Processing},
  volume    = {29},
  pages     = {3451--3460},
  year      = {2021}
}

@article{chen2022wavlm,
  author    = {Chen, Sanyuan and Wang, Chengyi and Chen, Zhengyang and Wu, Yu and Liu, Shujie and Chen, Zhuo and Li, Jinyu and Kanda, Naoyuki and Yoshioka, Takuya and Xiao, Xiong and Wu, Jian and Zhou, Long and Ren, Shuo and Qian, Yanmin and Qian, Yao and Wu, Jian and Zeng, Michael and Yu, Xiangzhan and Wei, Furu},
  title     = {{WavLM}: Large-Scale Self-Supervised Pre-Training for Full Stack Speech Processing},
  journal   = {IEEE Journal of Selected Topics in Signal Processing},
  volume    = {16},
  number    = {6},
  pages     = {1505--1518},
  year      = {2022}
}

@article{menne2024voice,
  author    = {Menne, Felix and D{\"o}rr, Felix and Schr{\"a}der, Julia and Tr{\"o}ger, Johannes and Habel, Ute and K{\"o}nig, Alexandra and Wagels, Lisa},
  title     = {The voice of depression: speech features as biomarkers for major depressive disorder},
  journal   = {BMC Psychiatry},
  volume    = {24},
  pages     = {794},
  year      = {2024}
}

@book{kim2012neurologic,
  author    = {Kim, Hyang-Hee},
  title     = {Neurologic Speech-Language Disorders},
  publisher = {Sigmapress},
  address   = {Seoul},
  year      = {2012}
}

@inproceedings{dumpala2025ttt,
  author    = {Dumpala, Sri Harsha and Sastry, Chandramouli S. and Uher, Rudolf and Oore, Sageev},
  title     = {Test-Time Training for Speech-based Depression Detection},
  booktitle = {Proc. Interspeech},
  year      = {2025},
  address   = {Rotterdam, Netherlands}
}

@inproceedings{gomezzaragoza2025foundation,
  author    = {Gomez-Zaragoza, Lucia and Marin-Morales, Javier and Alcaniz, Mariano and Soleymani, Mohammad},
  title     = {Speech and Text Foundation Models for Depression Detection: Cross-Task and Cross-Language Evaluation},
  booktitle = {Proc. Interspeech},
  year      = {2025},
  address   = {Rotterdam, Netherlands}
}

@inproceedings{zuo2025ordinal,
  author    = {Zuo, Lishi and Mak, Man-Wai},
  title     = {Leveraging Ordinal Information for Speech-based Depression Classification},
  booktitle = {Proc. Interspeech},
  year      = {2025},
  address   = {Rotterdam, Netherlands}
}

@inproceedings{you2025m3l,
  author    = {You, Jiajun and Wang, Shuai and Gong, Xun and Wan, Xiang},
  title     = {{M3L}: A Multi-Modal and Multi-Lingual Depression Detection Framework},
  booktitle = {Proc. Interspeech},
  year      = {2025},
  address   = {Rotterdam, Netherlands}
}

\end{document}